\documentclass[preprint,12pt]{elsarticle}
\usepackage{graphicx}
\usepackage{amssymb,amsmath}

\begin{document}

\begin{frontmatter}

\title{The topological issues of cooperation}
\author{M. N. Kuperman}
\address{Consejo Nacional de Investigaciones Cient{\'\i}ficas
y T{\'e}cnicas, Argentina} \address{Centro At{\'o}mico
Bariloche and Instituto Balseiro, 8400 S. C. de Bariloche,
Argentina }
\author{S. Risau-Gusman}
\address{Consejo Nacional de Investigaciones Cient{\'\i}ficas
y T{\'e}cnicas, Argentina} \address{Centro At{\'o}mico
Bariloche, 8400 S. C. de Bariloche, Argentina}

\begin{abstract}
\vspace{.5 cm}

In the last years the Prisoner Dilemma (PD) has become a paradigm
for the study of the emergence of cooperation in spatially
structured populations. Such structure is usually assumed to be
given by a graph. In general, the success of cooperative
strategies is associated with the possibility of forming globular
clusters, which in turn depends on a feature of the network that
is measured by its clustering coefficient. In this work we test
the dependence of the success of cooperation with the clustering
coefficient of the network, for several different families of
networks. We have found that this dependence is far from trivial.
Additionally, for both stochastic and deterministic dynamics we
have also found that there is a strong dependence on the initial
composition of the population. This hints at the existence of
several different mechanisms that could promote or hinder cluster
expansion. We have studied in detail some of these mechanisms by
concentrating on completely ordered networks (large clustering
coefficient) or completely random networks (vanishing clustering
coefficient).

\end{abstract}

\end{frontmatter}

\section {Introduction}

The emergence of cooperation in different real systems has been
puzzling researchers in several areas devoted to the study of
systems involving social, economic or biological organization.
Even though each of these systems is conformed by single units
with natural competitive tendencies, the emergence of collective
behaviours is undeniable. While natural selection operates through
competition, cooperation is essential to the evolution and
emergence of higher degrees of complexity. The struggle between
competition and cooperation is then one of the keys in
understanding the self organization of complex systems conformed
by interacting units. Still, many questions arise regarding how
such opposites forces can coexist.

The survival of the cooperative behavior is a classical problem of
game theoretical approaches \cite{mayn}. In this context, the
paradigmatic Prisoner's Dilemma game \cite{axel} has been widely
studied in different versions. It is usually formulated as a
standard model for the confrontation between cooperative and
selfish behaviors. For many years it was implemented in zero
dimensional systems, where every player can interact with any
other, until the crucial effects of spatial distribution were
finally noticed \cite {now0,now3}. Since then, several mechanisms
for the evolution of cooperation have been proposed. Some of them
are summarized in \cite{now1}: kin selection, direct reciprocity,
indirect reciprocity, group selection and network reciprocity.
Here we have chosen to focus on this last mechanism, which is
associated to the fact that a cooperative individual can take
advantage of the topology of the network to form clusters of
cooperators that are often resilient to the invasion of
cooperators.

Studies about the effect of network reciprocity have dominated the
literature on spatial distributed games in the last years
\cite{now3,kup1,kup2,sza2,lib,oht,roc}. They were the response to
the need of studying the evolution of the strategies of players of
a game beyond the simplifying assumption of a well-mixed
population, where everybody interacts equally likely with
everybody else. The observation that real populations are not well
mixed and the fact that spatial structures could affect the
evolution of a game and the strategies of the players demanded a
new approach. A natural step was to consider complex networks as
models for the underlying topology characterizing the spatial or
social structures. In the case of a game played on top of a
network or graph, the individuals of a population are located on
the vertices of the graph. The edges of the graph determine the
links through which individuals can interact. In a spatial model
for the prisoner dilemma, the players are classified either as
cooperators or defectors, and it is assumed that every agent can
only play with his/her neighbours.

It has been shown that extremely simple rules determine whether
network reciprocity can favor cooperation \cite{oht}. But it is
the concept that cooperators can prevail by forming clusters what
we want to analyze here. This idea has been discussed and analyzed
in many works. In \cite{now1} it is found that cooperators can
prevail by forming network clusters, where they help each other.
In one of the pioneering works on spatially extended games
\cite{now3} the authors have analyzed several shapes for a cluster
of cooperators and test the stability of each one against the
invasion by defectors. They found that cooperators can only
survive and grow if they form clusters. Another work pointing out
the clustering effect is \cite{doe} where it was stated that
cooperators can survive by forming clusters within which they
benefit from mutual cooperation, that in turn, allows them
screening the exploitation by defectors throughout the borders of
the cluster. It must be mentioned that some authors found an
inverse relationship between the formation of clusters and the
success of cooperation \cite{hauert}.

In this work we intend to show that the survival of cooperators
involves much more than the conformation of clusters. If it
depended only on that, cooperation success would increase
monotonically with the clustering coefficient of the underlying
network. By focusing on the analysis of networks which only differ
in their clustering coefficient, in the next sections we show not
only that this does not happen (i.e. the equilibrium  fraction of
cooperators is a non monotonic function of $C$) but also that
there is a strong dependence on the composition of the initial
population. This hints at the existence of several mechanisms
responsible for the expansion or extinction of cluster of
cooperators. To find these mechanisms, in the last section we
focus on what happens for populations in completely ordered and
completely disordered networks. A detailed analysis allows us to
understand the important of the initial fraction of cooperators
for the evolution of the different systems.

\section {The model}

The prisoner dilemma is a caricature of a real situation in which
selfish and altruist tendencies compete. It has been the subject of study of game theory for the last
60 years \cite{rapo,axel,now1}. Its name and formal elaboration is
attributed to A. Tucker, who mentioned it in a classroom in 1950, but
it was not until 1952 that the first results about it were published \cite{floo}.

The formulation of the prisoner dilemma as a game is rather
simple. It is played by two players who must choose their moves
between two strategies: to cooperate (C) or to defect (D). The
reward, or {\it payoff}, obtained by each player after one round
of the game is given by Table \ref{tabla1}:
\begin{table}[ht]
\centering
\begin{tabular}{c| c| c}
 &{\bf C} & {\bf D} \\
\hline
{\bf C} & $r$ & $s$  \\
\hline
{\bf D} & $t$ & $p$
\end{tabular}
\caption{Payoff table for the prisoner's dilemma: the strategy in
each row gets the payoff given by the table when playing again the
strategies in the columns.} \label{tabla1}
\end{table}

Each element in the payoff matrix represents the payoff of a
player using the strategies in the rows, when confronting a player
choosing the strategies in the columns. A defector ${\bf D}$
receives $t$, the temptation to defect, when its opponent is a
cooperator (${\bf C}$), who in turn gets $s$, the sucker's payoff.
In case of mutual cooperation, each player obtains a reward $r$,
while mutual defection punishes both players with the payoff $p$.
The table is in fact very general, because the payoffs of the
Prisoner's Dilemma must satisfy the additional constraints
$t>r>p>s$ and $2r>t+s$. Other relationships between the parameters
define the Snowdrift and Stag Hunt games \cite{sza2}.

In some versions of the game a different set of parameters is used: $r=c-b$, $s=-b$, $t=c$ and $p=0$
\cite{oht}, to account for a slightly different interpretation of
the game: a cooperator (C) is someone who pays a cost $c$ for any
other individual to receive a benefit $b$. In turn, a defector
does not distribute any benefits and gets those delivered by the
cooperators at no cost.

To simplify the analysis, in the following we use a reduced
version of the payoff table (Table 2), which has only one free
parameter. It has been show that this parameter eduction preserves
the most relevant features of the prisoner's dilemma~\cite{now0}.
\begin{table}[ht]
\centering
\begin{tabular}{c| c| c}
 &{\bf C} & {\bf D} \\
\hline
{\bf C} & $1$ & $1-t$  \\
\hline
{\bf D} & $t$ & $0$  \\
\end{tabular}
\caption{Reduced payoff table for the prisoner's dilemma: the
strategy in each row gets the payoff given by the table when
playing again the strategies in the columns.} \label{tabla2}
\end{table}

In order to study the possibility that the players can change
their strategies as a result of their previous interactions, thus
generating an evolutionary dynamics of strategies, many authors
started to work with the iterated Prisoner Dilemma, in which
players interact by iteratively playing the game several times.
The history of successes or failures of each player is recorded in
what is called his cumulative payoff. How the players use the
information accumulated in their own and others cumulative payoffs
is what defines the rules of evolution. Operationally, the
evolutionary dynamics acts at a certain instance of the game, for
example after everybody has played against everybody else, when
players decide whether to change strategies or not, following
certain update rules. Before all the players start again playing
the game, all the cumulative payoffs are set to $0$. The spectra
of rules of evolution is wide and ranges from purely deterministic
to stochastic dynamics \cite{oht,sza,now2,moy,hau}.

Complementary to the evolutionary aspects mentioned above, many
authors started to analyze spatial games in order to cope with the
limitations associated with the assumption that players were
always part of a well mixed population. \cite{now0,now3,sza2}.

The evolutionary behaviour of the populations of surviving
strategies of spatial games on networks can be affected by several
features of the underlying topology as, for example, the degree
distribution of the graph, the average distance between nodes, or
the clustering coefficient \cite{kup1,kup2,dur,moy,roc}.

The concept that cooperators can survive by grouping in clusters
has been discussed and analyzed in many works \cite{oht,now3,doe}.
Intuitively, the reasoning goes as follows. The effect of the
cluster would be to screen the nodes at the interior from the
presence of defectors. As defectors can only get an advantage from
their interaction with cooperators, only those located next to the
border of a cluster of cooperators should collect any benefits. In
turn, although the cooperators at the border of the cluster should
have lower payoffs because of their interaction with defectors,
their cooperator neighbours at the interior of the cluster should
perform better than the defectors at the border. Thus, imitating
the internal cooperators should be always more convenient than
imitating the bordering defectors, which should lead to the
survival, and even expansion, of the cluster of cooperators. The
problem is that all these arguments, as well as the very
definition of `cluster', depend crucially on the structure of the
network. The most important feature in this regard is the
clustering coefficient $C$, which measures how connected is the
neighbourhood of each node, on average. The existence of local
transitive relationships, closely related to the clustering
\cite{was}, is what defines the possibility of survival of small
clusters of cooperators. Paradoxically, it will be also
responsible for the negative effect that an isolated cooperator
may have on incipient cooperative clusters.

Here we use the definition of global clustering coefficient of
Watts and Strogatz~\cite{watts}. For each node $i$, its local
clustering coefficient is defined as the quotient between the
number of links joining nodes of the neighbourhood of $i$ divided
by the total number of possible links ($k_i(k_i+1)$). $C$ is then
defined as the average over $i$ of all local clustering
coefficients. We study the influence of $C$ on the evolutionary
dynamics of the iterated prisoner's dilemma, but keeping the
degree distribution constant, to disentangle both contributions.
For this we analyze regular networks (i.e. with the same number of
neighbours for every node) with different values of $C$, generated
with the following algorithm. Starting from an ordered network
(defined below) we select at random two pairs of connected nodes.
Then we `cut' both connections and connect each individual to one
of the individuals it had not been connected before. In other
words, the connections are swapped. If this change gives a network
with larger $C$, it is accepted and the network is updated. If it
does not increase $C$, the change is only accepted with a fixed
(and typically small) probability. This process goes on until the
clustering coefficient has reached the desired value. Notice that
this procedure leaves the degree distribution of the original
network unchanged. When the desired clustering coefficient is very
low, it is to be expected that the resulting networks is very
close to a regular random network, independently of the starting
one. On the other hand, for larger values of $C$ it is to be
expected that the effect of the starting network is much larger.
For this reason we use two different starting networks: ring
networks where each node is connected symmetrically to the closest
$k$ nodes, and 2-dimensional lattice networks. The networks
generated from these two classes are called, respectively, random
ring networks or random lattice networks. Three different starting
lattice networks are used: regular square lattices ($k=4$),
triangular lattices ($k=6$) and square lattices where each node is
connected to its Moore neighbourhood ($k=8$). For all values of
$k$ ring networks can be considered as one-dimensional because for
a given cluster of nodes the size of the surface is independent of
the volume whereas for lattice network the relationship is $V
\approx S^2$.

Throughout our simulations, we have considered two types of
evolutionary dynamics, one deterministic \cite{sza2} and the other
stochastic~\cite{oht}. In both cases, each player either copies
the strategy of one of its neighbours or sticks to the same
strategy used in the previous round. In the deterministic dynamics
each player copies the strategy of its most successful neighbour,
if the payoff of that neighbour is larger than its own. In the
probabilistic dynamics, previously used in \cite{oht}, it copies
the strategy of a neighbour chosen at random, with a probability
proportional to its relative payoff. His own strategy is also
included in the pool of eligible strategies. As the results we
have obtained are qualitatively the same for both types of
dynamics, in the following we focus on the deterministic dynamics,
and comment briefly on the small differences obtained when using
the stochastic dynamics

\section {Numerical Results}

As mentioned in the previous section, we consider two different
dynamics, though explicit results corresponding to only one of
them will be shown in the following paragraphs. In all the cases
we consider regular networks with 1000 to 10000 nodes with even
degrees between 4 and 8. We observe no dependence on the size but
different regimes associated to the degree. The state of the nodes
is synchronically updated and the payoff of each player is not
cumulative in time. Even though we observe that different initial
concentrations of cooperators, $\rho_c(0)$, lead to qualitatively
the same results, when properly scaled, there are some important
differences. To show this we use two different initial
concentrations of cooperators, $\rho_c(0)=0.1$ and $\rho_c(0)=0.5$
for every network analyzed in this paper.

\begin{figure}
\centerline{\includegraphics[clip=true]{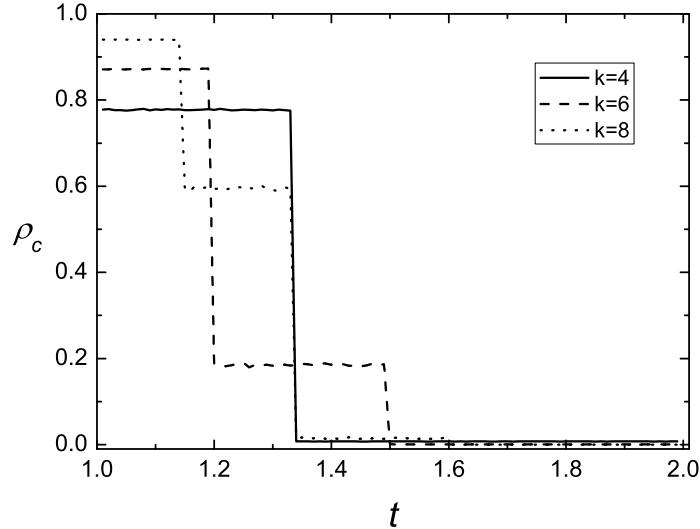}} \caption{
Steady cooperator density $\rho_c$ as a function of the parameter
$t$ for three different networks: $k=4$ (lattice), random lattice
with $C=0.2$ starting from a lattice with $k=6$, and idem with
$k=8$.} \label{figsteps}
\end{figure}

If the equilibrium value of $\rho_c$ is plotted as a function of
$t$, leaving all the other parameters constant, a piecewise
constant function is obtained as is shown in Fig.1. This has also
been previously noticed \cite{dur}, but with a different payoff
table (in the case considered in \cite{dur} a cooperator gets $0$
payoff when playing against a defector). To understand the origin,
and quantify the limits, of these steps, we must consider the
necessary conditions for the propagation of the cooperating
behavior. For a cooperator to have a chance to turn a defecting
neighbour into a cooperating one, its payoff should be at least
larger than that of the defecting neighbour. This leads to the
condition $n_{CC}+(k-n_{CC})(1-t) > n_{DC} t$, where $n_{CC}$ is
the number of cooperator neighbors of the cooperator and $n_{DC}$
is the number of cooperator neighbors of the defector. The
condition on $t$ can be written as $t>k/(k-n)$ where
$n=n_{CC}-n_{DC}$. Note that, as $n_{CC} \leq k-1$ and $n_{DC}
\geq 1$, $n$ is a natural number that must satisfy $1 \leq n \leq
k-2$. This gives a maximum of $k-1$ possible steps. Note however
that in some networks the range of possible values for $n$ is
smaller, and therefore the number of steps of $\rho_c$ is at most
$k-2$. In general, for networks with the same number of $k$ the
number of possible steps will be smaller for the networks with
smaller clustering coefficients. As an example, consider the two
extreme cases of a tree and a lattice network with $k=8$: whereas
the tree has the maximum possible of steps, the lattice network
can have at most $4$ steps. In all cases the last step corresponds
to $\rho_c=0$ because for those values of $t$ a cooperator,
regardless of the composition of its neighbourhood, is not able to
turn a defecting neighbour into a cooperating one. Furthermore, it
is also possible that, because of geometrical constraints,
$\rho_c$ also vanishes for other steps. For the networks analyzed
in this paper, we have confirmed that only the height of the steps
depends on $C$. Furthermore, simulations show that only for the
first two steps the final number of cooperators is non vanishing
(see Fig.\ref{figsteps}). For these reasons we have only analyzed
the dependence of $rho_c$ in these first two steps, i.e. we have
used only two values of $t$, $t_1$ and $t_2$, that satisfy
$1<t_1<k/(k-1)$ and $k/(k-1)<t_2<k/(k-2)$.

\begin{figure}
\centerline{\includegraphics[clip=true]{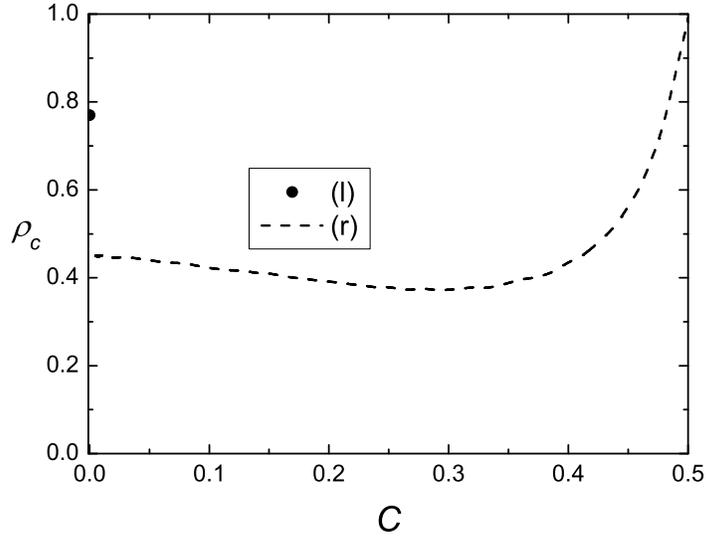}} \caption{
Steady cooperator density $\rho_c$ as a function of the clustering
coefficient $C$, for $k=4$ and $\rho_c(0)=0.5$. (l) and (r) in the
caption refer to lattice and ring networks respectively.}
\label{k4}
\end{figure}

\begin{figure}
\centerline{\includegraphics[clip=true]{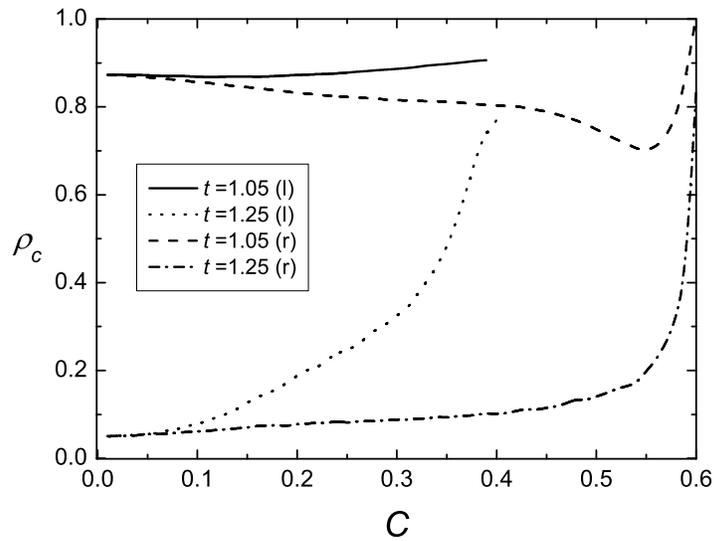}} \caption{
Steady cooperator density $\rho_c$ as a function of the clustering
coefficient $C$, for $K=6$ and $\rho_c(0)=0.5$. (l) and (r) in the
caption refer to lattice and ring networks respectively.}
\label{k6}
\end{figure}

\begin{figure}
\centerline{\includegraphics[clip=true]{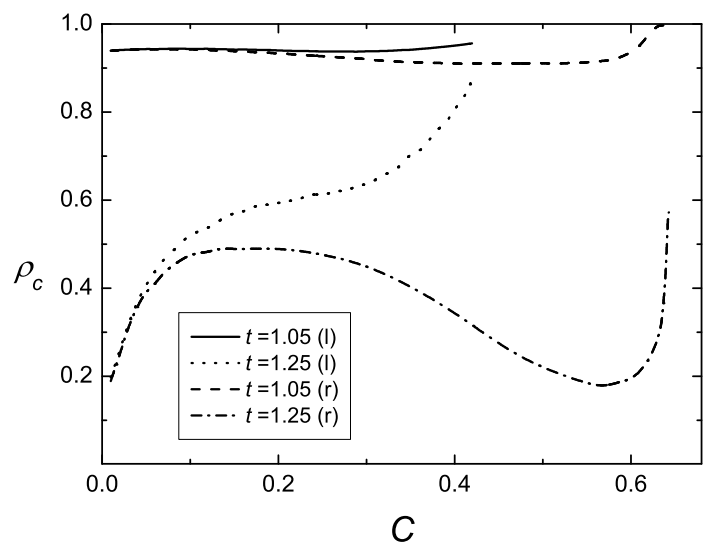}}
\caption{Steady cooperator density $\rho_c$ as a function of the
clustering coefficient $C$, for $K=8$ and $\rho_c(0)=0.5$. (l) and
(r) in the caption refer to lattice and ring networks
respectively.} \label{k8}
\end{figure}

In Figs.~\ref{k4} to~\ref{k8} we plot the numerical results
obtained from computational simulations with $1000$ to $5000$
agents. Each curve corresponds to the average fraction of
cooperators in the steady state, as a function of the clustering
of the networks. The highest clustering value corresponds to the
ordered network (lattice or ring), and networks get increasingly
disordered as $C$ is decreased.

We begin by analyzing what happens for evolutions whose initial
state consists of the same number of cooperators and defectors. In
other words, the initial probability that a given agent is a
cooperator is $0.5$. In this case, the steady state is always
composed by a finite fraction of cooperators. As can be seen in
the figures, there are some features that are common to all the
families of networks analyzed. The first is that, for each class
of network, the behaviors of the curves is qualitatively the same
for the two values of $t$ used. The only difference is that, as is
to be expected, curves for $t_1$ are below curves for $t_2$.
Another important feature is that the final fraction of
cooperators for ordered networks is always larger than what is
obtained in completely random networks. Even though this seems to
confirm the impression that clustering is beneficial to
cooperators, it must be noticed that many curves are not monotonic
with $C$, as for example, all curves corresponding to random ring
networks.

Another interesting feature to notice is that, for ring networks
the addition of a very small amount of disorder causes an abrupt
decrease in the steady  fraction of cooperators. This happens
because of the one-dimensional nature of the ring: rewiring very
few links at each side of a cooperation cluster can be very
effective in stopping its expansion. When more links are rewired
the dimensionality of the system begins to increase and
cooperators clusters find new directions to expand.

For all values of $C$ curves for random ring networks are  always
below those for random lattice networks, for the same values of
$t$. This is probably related to the lower dimensionality of the
substrate of the random ring network that may have an influence
even for high values of the disorder. Notice that the curves only
overlap for very small values of the clustering coefficient. This
means that a large amount of disorder is needed for the network to
`forget' the starting substrate.

In Fig.~\ref{k4} only one point is shown for random lattice
networks because both the square lattice and the completely random
network with $k=4$ have a vanishing clustering coefficient. The
large difference seen in Fig.~\ref{k4} between the steady state
fraction of cooperators could be attributed to the much shorter
minimal distances between nodes in random regular networks (which
have a diameter $\approx \log N$~\cite{bollo}) or to the presence
of short loops in the square lattice (see next section).

\begin{figure}
\centerline{\includegraphics[clip=true]{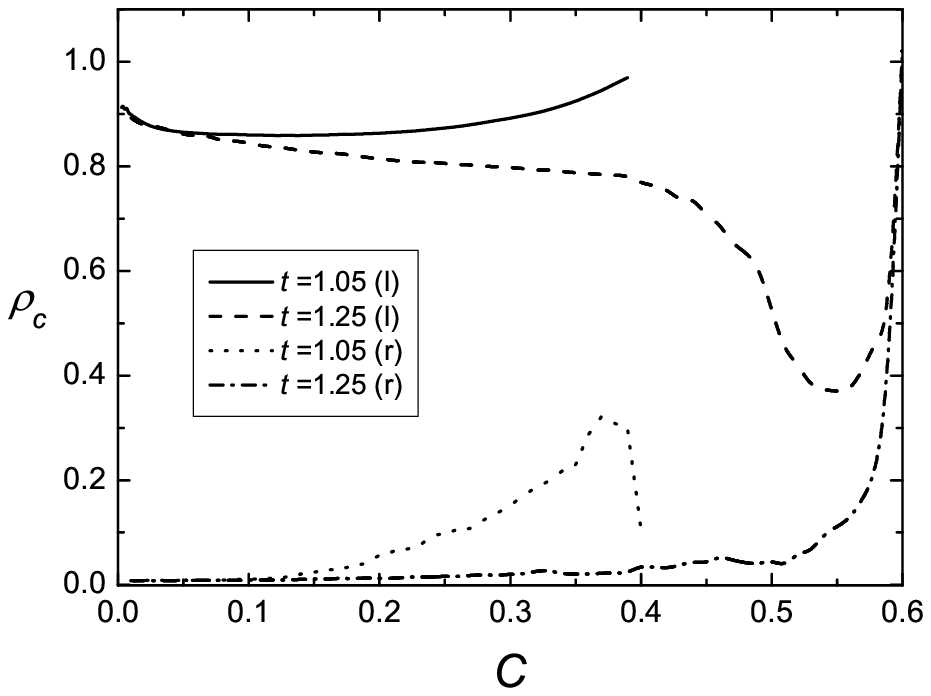}} \caption{
Steady cooperator density $\rho_c$ as a function of the clustering
coefficient $C$, for $K=6$ and $\rho_c(0)=0.1$. (l) and (r) in the
caption refer to lattice and ring networks respectively. Only
those realizations that showed to the survival of cooperators were
considered. } \label{k6b}
\end{figure}

When the initial state has less collaborators, the situation is
more complex to analyze because for some systems the population
evolves to an equilibrium state where all the cooperators have
been eliminated. If, however, we consider only those systems that
have a steady state with a non vanishing fraction  of cooperators,
the picture is very similar to what is found for $\rho=0.5$. An
example of this for $k=6$ is shown in Fig.~\ref{k6c} where the
initial fraction of cooperators was $\rho_c=0.1$ (compare
Figs.~\ref{k6b} and~\ref{k6c}).

\begin{figure}
\centerline{\includegraphics[clip=true]{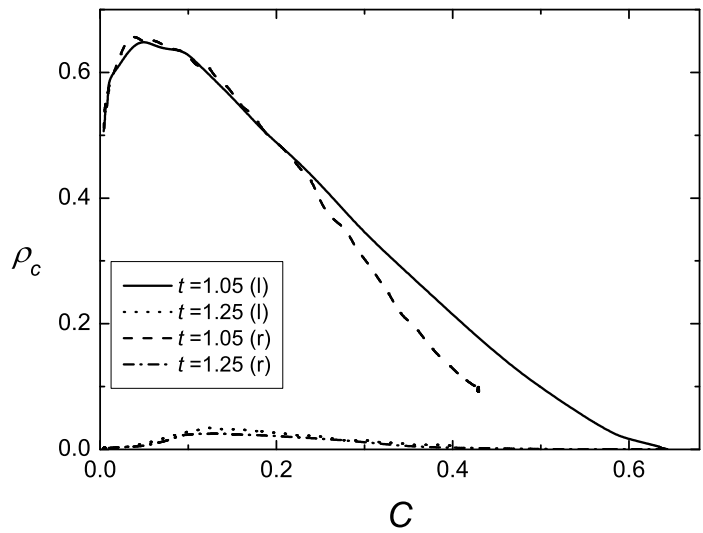}} \caption{
Fraction of realizations that converge to a steady state with a
positive number of cooperators, for networks with $k=8$. }
\label{k8d}
\end{figure}

When, instead, the fraction of realizations that converge is
considered, the picture that emerges is rather different, as
Figs.~\ref{k8d} and ~\ref{k6c} show. In this case, ordered
networks are less favourable for the preservation (and eventual
expansion) of cooperation than completely random networks. As
before, the behaviour between these two extremes is not monotonic.
A feature of these curves that stands out is that, for the same
values of $C$, they seem to depend very weakly on the type of
substrate used to generate them.

\begin{figure}
\centerline{\includegraphics[clip=true]{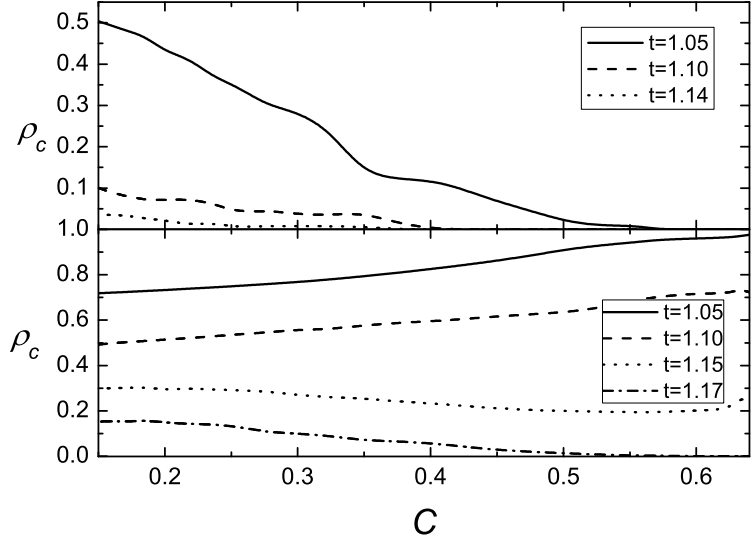}} \caption{
Average value of the steady-state fraction of cooperators for a
stochastic evolutionary dynamics, for several values of $t$ for
random ring networks with $k=8$, and for $\rho_c(0)=0.5$ (panel A)
and $\rho_c(0)=0.1$ (panel B).} \label{figstoch}
\end{figure}

\begin{figure}
\centerline{\includegraphics[clip=true]{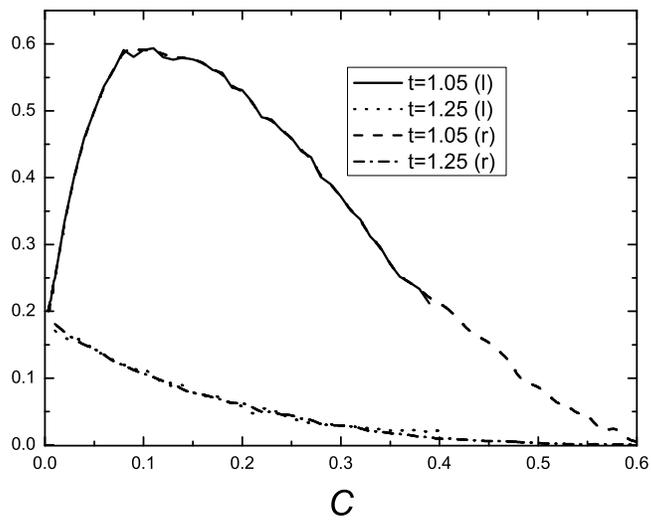}} \caption{
Fraction of realizations that converge to a steady state with a
positive number of cooperators, for networks with $k=6$. }
\label{k6c}
\end{figure}

The difference of the evolutionary dynamics of populations
starting from many, or few, cooperators can be shown even more
clearly using the same variable for both cases: the average
fraction of steady state cooperators, with the average taken over
the whole population. But it must be recalled that for populations
with small numbers of initial cooperators the variable does not
give atypical value of final cooperators because the steady state
cooperators distribution has at least two modes clearly separated,
one with zero cooperators and other with many cooperators. We have
used this variable to show that for some stochastic dynamics the
results are very similar to what has been described above for a
deterministic dynamics. In the stochastic dynamics we have used,
the agents choose the strategy of a neighbor with a probability
proportional to the corresponding cumulative payoff, but only if
it is larger than his/her own cumulative payoff~\cite{oht}.
Fig.~\ref{figstoch} shows some results for this dynamics. Curves
for several different values of $t$ are shown because in this
case, the dependence on $t$ is not as simple as in the
deterministic case. In any case, several qualitative similarities
with the deterministic case are apparent. For $\rho=0.5$ the
ordered networks are more favourable to cooperation than
completely disordered ones, at least for $t \leq 1.15$. For
$\rho=0.1$ the situation is reversed, and now the most favorable
networks in terms of cooperation are completely random ones. There
are even some values of $t$ for which the dependence with $C$ is
not monotonic.

So far we have shown that the clustering coefficient of the
networks has an important influence on the steady state. But
whether or not this influence is beneficial depends strongly on
the initial fraction of cooperators, both for deterministic and
stochastic evolutionary dynamics. The possible causes for this are
addressed in the next section.

\section {Role of the initial fraction of cooperators}
To understand the conflicting ways in which network clustering can
affect cooperation we study in some detail the deterministic
dynamics, and hope that some conclusions apply also to the
stochastic case. Furthermore, in terms of the range of possible
clustering coefficients we limit ourselves to analyze what happens
for networks at the two extremes: completely ordered (large $C$)
and completely disordered (small $C$) networks. For this last
class we concentrate on lattice networks with $k=6$ and $k=8$
which have clustering coefficients $C=2/5$ ($k=6$) and $C=3/7$
($k=8$),

First, we analyze the fate of a cluster of 3 cooperators. In the
case of the lattice networks  with $k=6$ and $k=8$, there are two
and three possible configurations, respectively (see
Fig.\ref{k8t}). However, all of them are unstable because, thanks
to the large clustering, some neighbors of the cluster can be
connected to 2 or 3 cooperators in the cluster, having thus a
larger payoff than any of them. On the other hand, in a random
network there is a finite probability ($1-3k^2/N+O(k/N)$) that all
the neighbors of a 3-cluster are not neighbors of more than 1
cooperator, and thus the cluster is stable. Furthermore, if this
cluster does not disappear it will grow to become a cluster of
$k+1$ cooperators (a central cooperator surrounded by
cooperators). This cluster, in turn, has a non vanishing
probability of continuing its expansion. For example, if there is
a link joining two of the new surrounding cooperators (which
happens with probability $(k/N) k(k-1)/2)$, the cluster grows by
turning into cooperators the $2(k-2)$ non-cooperating neighbors of
the nodes that share the link (see Fig.~\ref{figer}A). But there
is now a non vanishing probability that there is also a link
joining the nodes of the cluster 'surface' which would lead to an
increase of its size of $2(k-2)$. And, in general, at any step of
its growth it could keep growing with a probability roughly
proportional to $(k/N) N_s 2(k-2)$, where $N_s$ is the size of the
cluster surface. Considering that most of the nodes of the
clusters lie in its surface, implies that once the cluster has
reached a size of order $N/(2k(k-2))$ it will keep growing until
it spans the whole lattice. Thus, if we consider the evolution of
all possible 3-clusters, the distribution of final cluster sizes
should be non-vanishing only for sizes $O(N)$ and for sizes
smaller than $O(N/(2k(k-2)))$.

In the case of 4-clusters, it is easy to see that the situation is
qualitatively the same for regular random networks. On the other
hand, for lattice networks the picture is completely different:
square 4-clusters expand until they occupy the whole lattice
because every cooperator is connected to 2 or 3 others whereas
non cooperating neighbours can only have at most 2 cooperating neighbours.

Taking these ideas into account, and assuming that cluster
expansion or death is not influenced by the presence of
cooperators outside the cluster, we can try to predict what
happens when cooperators are placed at random in a network. In the
case of a lattice network, the probability of an initial set of
cooperators taking over the network is simply the probability that
there is at least one square cluster of cooperators:
$P(p)=1-(1-p^4)^N$. For a random regular network the probability
is $P(p)=1-(1-f(p,k,N))^N$, where $f(p,k,N)$ is the probability
that a given node is the center of a 3-cluster and that it expands
during at least two steps:

\begin{eqnarray}
f(p,k,N)  &=& (1-(1-k/N)^{4(k-2)(k-2)})(1-(1-k/N)^{\frac{k(k-1)}{2}}) \nonumber \\
&& (1-(1-p)^k-kp(1-p)^{k-1})
\end{eqnarray}

\noindent where the first term in the product is the probability
that a given node has at least two cooperating neighbors, the
second is the probability that the 3-cluster expands in the first
step, and the third is the probability that it continues expanding
in the second step. We assume that after the second step the
expansion goes on until all the lattice is occupied by
cooperators. Fig.~\ref{k8t} shows that these functions overestimate
the fraction of populations that are able to take over the whole
network. This shows that, somewhat paradoxically, the presence of
other cooperators can sometimes hinder the expansion of a cluster.
For example, if a square has 2 defecting neighbors, connected to
opposite sides of the square, and in turn connected to at least
one cooperator outside of the cluster, the square disappears. Note
that a 4-cluster in a random network is harder to destroy because
a neighbour of the cluster is a neighbour only to one cooperator
of the cluster and thus it needs two other cooperating neighbors
to be able to destabilize the cluster. Evidently, 3-clusters are
much easier to destabilize, as the partial failure of the estimate
shows. Fig.~\ref{figer}B shows one way a 3-cluster can be destabilized by a close cooperator.

Interestingly, there is also a mechanism by which
different clusters can collaborate in each other's expansion,
given that the average distance between nodes is small enough.
Consider for instance a couple of stable stars of cooperators in a
random network. If they are connected, i.e. if a cooperator of one
star is connected to a cooperator in the other, these cooperators
would be able to turn their non-cooperating neighbors into
cooperators, thus increasing the size of the two clusters by
$(2k-2)$. In a random network, the probability that two clusters
of size $N_1$ and $N_2$ are connected is $1-(1-k/n)^{N_1 n_2}$.

\begin{figure}
\centerline{\includegraphics[width=10cm, clip=true]{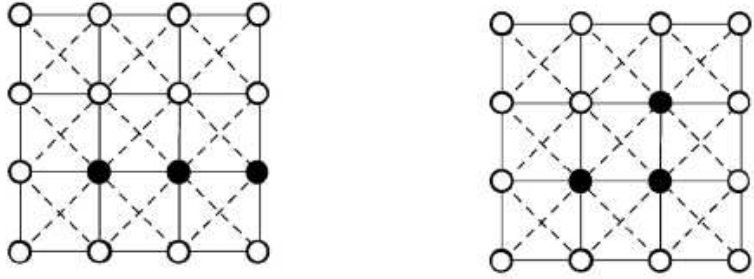}}
\caption{3-clusters in a lattice network with $k=8$. Black circles
represent cooperators and white circles represent defectors.}
\label{figmoore}
\end{figure}

Taking all these features into account it is possible to give a
better estimate for the fraction of systems that converge to a
final state dominated by cooperators or with only a few ($O(1)$)
stable cooperators. The probability that the cooperator population
dies out is $(1-p P_s)^N$ where $P_s$ is the probability that a
cooperator survives the first time step:
\begin{eqnarray}
P_s(k,P,N) &=& \sum_{j=2}^k \binom{k}{j} p^j (1-p)^{k-j} \nonumber \\
&& \left( \sum_{i=0}^{j-2} \binom{k-1}{i} p^i (1-p)^{k-1-i}
\right)^{k-j}
\end{eqnarray}

\begin{figure}
\centerline{\includegraphics[width=10cm, clip=true]{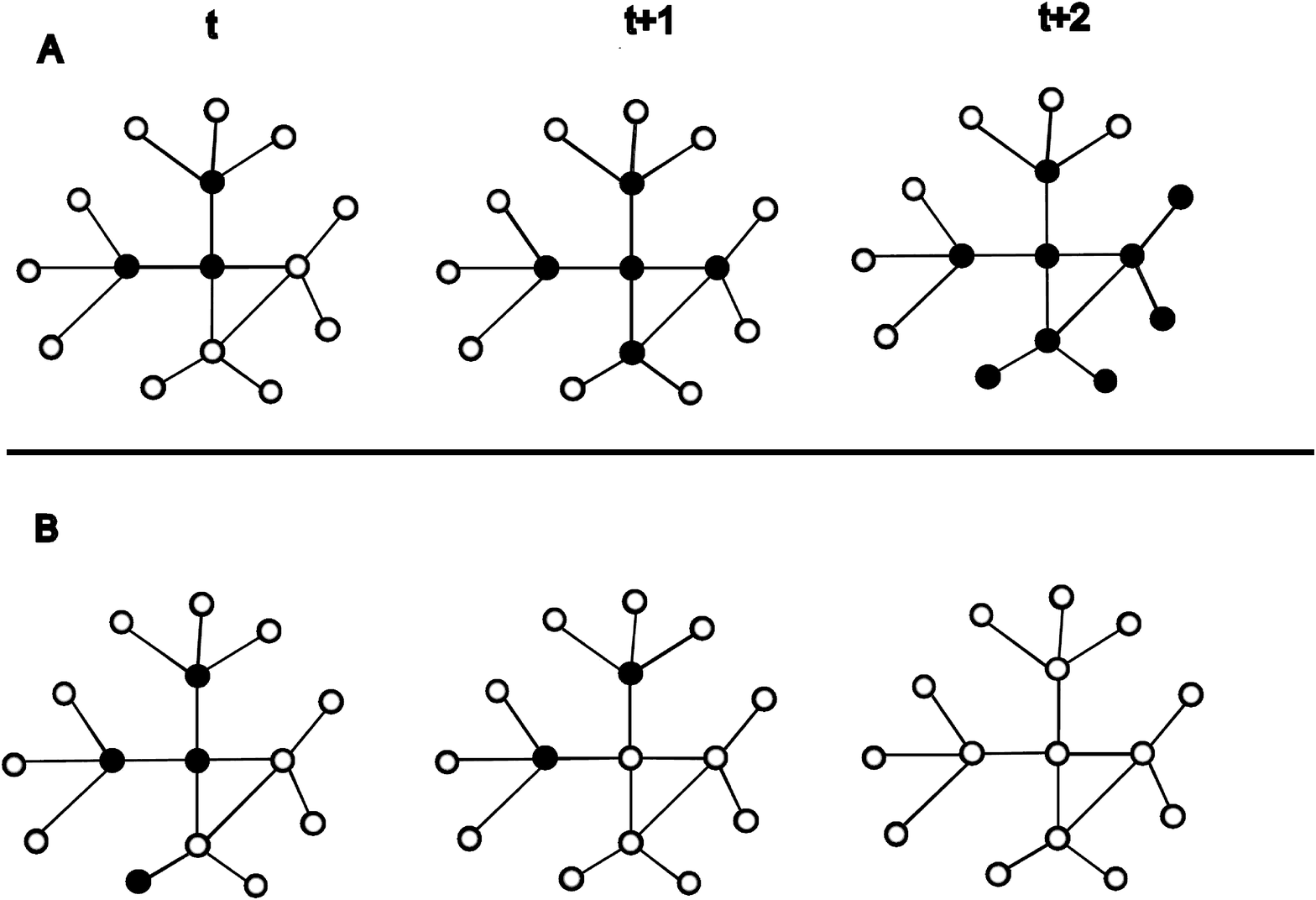}}
\caption{Evolution of two different 3-clusters in a random network
with $k=4$. Black circles represent cooperators and white circles
represent defectors. A) Expansion of a 3-cluster. B) Disappearance
of a 3-cluster.} \label{figer}
\end{figure}

\noindent which is simply the sum of the probabilities of having
$j$ cooperator neighbours multiplied by the probability that none
of the $k-j$ defecting neighbors has more than $j-2$ cooperating
neighbors. The probability of the final state being dominated by
cooperators can be approximated by:
\begin{eqnarray}
P(p) &=& 1-(1-P_s)^{pN}- \nonumber \\
&& \sum_{i=1}^{pN} \binom{pN}{i} P_s^i (1-P_s)^{pN-i}(1-k/N)^{k
i(k i-1)/2}
\end{eqnarray}

\noindent Each addend gives the probability of having $i$ stars
and that they are not connected.

\begin{figure}
\centerline{\includegraphics[clip=true]{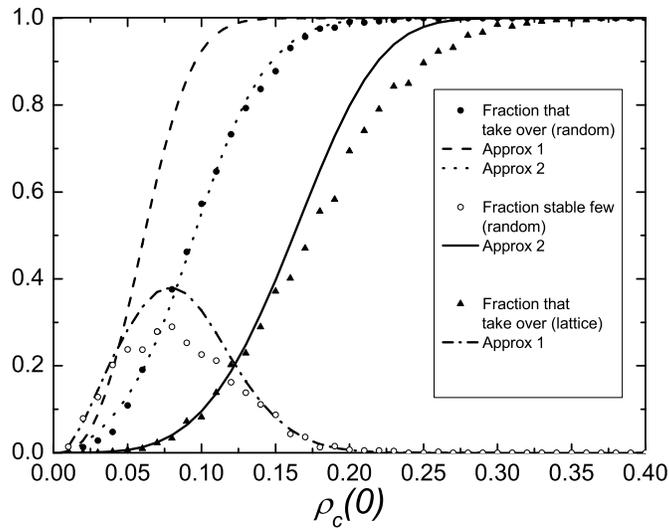}} \caption{
Fraction of systems that converge to a state dominated by
cooperators (full symbols) or to a state with a few stable
cooperators (empty symbols), as a function of the initial fraction
of cooperators, for regular random networks (circles) and lattice
networks (triangles), with $k=8$. The lines show the theoretical
estimates, assuming independence (full lines) or dependence
(dashed lines) among all clusters of cooperators.
 }
\label{k8t}
\end{figure}

\begin{figure}
\centerline{\includegraphics[clip=true]{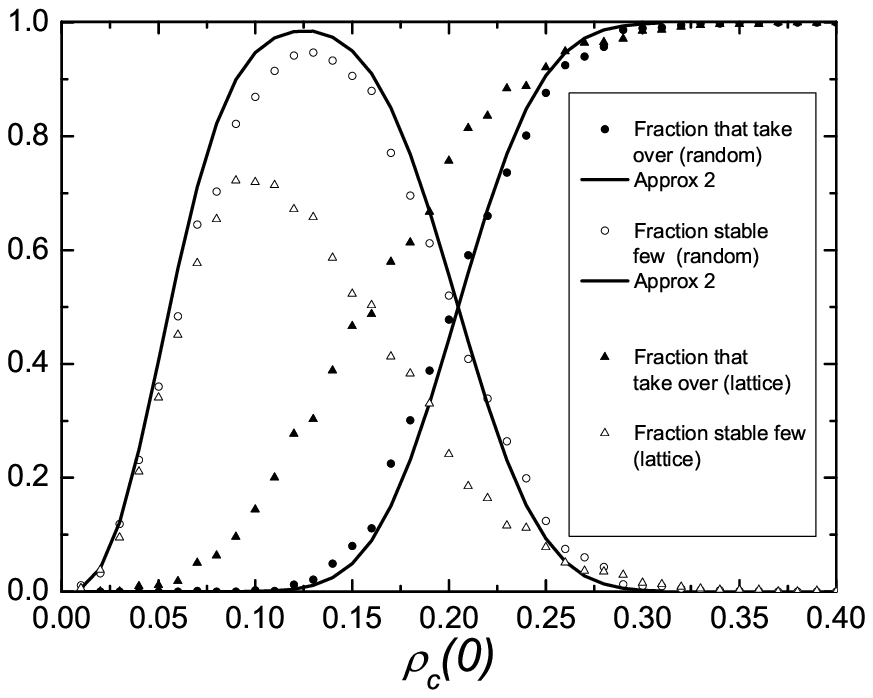}} \caption{
Fraction of systems that converge to a state dominated by
cooperators (full symbols) or to a state with a few stable
cooperators (empty symbols), as a function of the initial fraction
of cooperators, for random regular networks (circles) and lattice
networks (triangles), with $k=4$. The lines show the theoretical
estimates assuming dependence among all clusters of cooperators. }
\label{k4t}
\end{figure}

It is interesting to see what happens for networks with $k=4$
because both lattice and random regular networks have vanishing clustering
coefficients. Fig.~\ref{k4t} shows that the probability of
taking over the whole population is rather similar for both
networks. Interestingly, in this case there is also the
possibility of having a final state with a small number of stable
cooperators, for the lattice network. The reason is very similar to
the case of random regular networks and is a consequence of having a vanishing
cluster coefficient: a lineal cluster of $3$ cooperators cannot be
destabilized because no neighbor can be a neighbor to more than
$1$ cooperator of the cluster.

\section {Conclusions}

It has been sometimes suggested that one of the possible reasons
for the success of cooperating strategies in spatially structured
populations is the possibility of forming globular clusters. In
this way, cooperators inside the cluster are `protected' by the
ones on the border. If the populations is placed on a graph, the
`globularity' of the possible clusters is proportional to the
clustering coefficient. Therefore, cooperating strategies should
be more successful in networks with large $C$ than in networks
with small $C$. For the evolutionary dynamics studied here we
confirm that this is indeed the case when the fraction of
cooperators in the steady state is compared in random regular
networks (low $C$) and lattice networks (large $C$) having the
same degree distributions. However, by analyzing populations in
graphs with intermediate values of $C$, we find that the
equilibrium fraction of cooperators is not a monotonic function of
$C$.

The results commented above were obtained using initial
populations with many cooperators (half of the population, on
average). But if the initial population has much less cooperators,
the situation becomes more complex. On the one hand one finds that
some evolutions lead to the elimination of all cooperators. On the
other hand, when cooperators do not disappear, their final
fraction, as a function of $C$ has a similar behaviour to that
observed when there are many initial cooperators. The problem is
that the number of such evolutions is much smaller for ordered
networks than for disordered ones, the behaviour for intermediate
values of $C$ also being non monotonic. The situation is then very
different from that obtained from initial populations with more
cooperators.  The same difference appears when stochastic
evolutionary dynamics are analyzed.

The non monotonicity of the curves, together with the dependence
on the initial condition suggest that there might be several
mechanisms that influence the success or failure of cooperation.
In the last section we have shown that this is indeed the case, at
least for completely ordered or completely disordered networks. In
ordered networks the evolution is isotropic and deterministic:
wherever it is placed, a square cluster of 4 cooperators is always
able to expand. But in disordered networks the fate of the cluster
depends on where in the network it is located: given a large
enough network there are positions from where a cluster of 3
cooperators will be able to grow to a very large size. In other
words, in disordered networks smaller clusters are able to expand
than in the case of ordered networks, but only if they are placed
in the right places. Additionally, we have shown that it is not
uncommon that cooperators that are outside, but not very far, from
a cluster of cooperators, can actually hinder its expansion and
even lead to its disappearance.

\end{document}